# Current Problems of Magnetic Storm Prediction and Possible Ways of Their Solving

## O.V. Khabarova

Space Research Institute, Russian Academy of Sciences, Moscow, Russia

e-mail: olik3110@list.ru

Main problems of magnetic storm prediction and causes of low efficiency of medium-term prognosis are discussed. It is supposed, that possible way of their solving is searching for poor-investigated features of solar wind (for instance, solar wind density behavior before storms). The necessity of investigation not only severe storms and coronal mass ejections (CMEs), but origin of moderate and weak storms is shown. Examples of successful investigations in this direction are given.

## Introduction

Successful magnetic storm forecasting is one of the main aims of the space weather investigations. Forecasting methods can be classified into short-term (about 1 *hour* in advance, using spacecraft data), medium-term (about 1-4 *days*), and long-term (>7 *days*, solar cycle intensity predictions). Features of prognostic models and their accuracy depend on the alert time ΔT (see Table 1).

**TABLE 1**

**Most popular on-line web-pages on geomagnetic storm prognosis**

|  | http-address | Advance time ΔT |
|---|---|---|
|  | **Long- and medium-term forecasts** |  |
| Institute of applied geophysics, Moscow | http://www.meteorf.ru/srv/ipg/ipg_home.htm ; http://www.geospace.ru | 7 days |
| Institute of terrestrial magnetism, ionosphere and radio-wave propagation (IZMIRAN), Moscow | http://www.izmiran.rssi.ru/space/solar/forecast.shtml ; http://www.izmiran.rssi.ru/~romash/ ; http://forecast.izmiran.rssi.ru/prognoz/progn.html | 1÷7 days |
| ISES Regional Warning Centre for Canada | http://www.spaceweather.gc.ca/forecastmap_e.shtml | 1 day |
| The Space Weather Bureau (NASA) | http://www.spaceweather.com | 2 days |
| Australian Government, IPS Radio and Space Services | http://www.ips.gov.au/Main.php?CatID=4&SecID=3&SecName=Summary%20and%20Forecasts&SubSecID=1&SubSecName=Daily%20Report | 3 day |
| University of Lethbridge | http://www.spacew.com | 1-3 days |
| Naval Research Laboratory | http://wwwppd.nrl.navy.mil/whatsnew/prediction/index.html | ≤ 1 day |
|  | **Short-term forecasts** |  |
| Space Research Institute, Russia | http://iki.cosmos.ru/apetruko/forecast | ~ 1 hour |
| Solar-Terrestrial Physics Division, Danish Meteorological Institute, Denmark | http://dmiweb.dmi.dk/fsweb/solar-terrestrial/staff/wu/spwrtpdst.html | ~ 1 hour |
| Regional Warning Center, Sweden (RWC-Sweden) | http://www.lund.irf.se/rwc | ~ 1 hour |

The short-term forecasts are based on the information from spacecraft in the Sun-Earth libration point and different statistical models, connecting near-Earth plasma conditions with the geomagnetic field disturbance level (the fact that electromagnetic signal from spacecraft propagates faster than plasma is used here). Such forecasts are rather exact, up to ~90%, but their alert time (ΔT~1 *hour*) is too small for preventing of storm hazard (see references at corresponding web-sites, given in Table 1).

The long-term forecasts try to predict general space weather and geomagnetic situation in relative far future, using solar observations and different statistical models. There is no correct information about the accuracy level of this type of forecasts, and they are usually used for the academic interest only.

The medium-term forecasts are most valuable for practical aims. Methods of their realization are mainly based on the recognition of approach of geoeffective structures to the Earth. Since interplanetary coronal mass ejection (ICME) interactions with the Earth's magnetosphere are considered as cause of super-intense geomagnetic storms [1-5], investigations of CME's features are carrying out incessantly, and CME-like conditions in solar wind are taken for prognostic aims as geoeffective by default for the all type of magnetic storms. Most of the medium-term forecast methods are oriented towards the prediction of probability of severe storms with *Dst* < - 80 *nT* only (see references in Table 1).

Meanwhile, the medium-term forecasts' quality remains rather modest: even during solar maximum, when the number of CMEs is large, the successful forecasting rate is ~ 75% (see, for example, Naval Research Laboratory's web-page). This level falls down during solar minimum [6], because CMEs are rare during this period.

But actual forecast quality is lower, because the number of geomagnetic storms with *Dst* < - 80 *nT* is always less than 10% of the total [2].

## Main problems of magnetic storm prediction

Therefore the main problem of medium-term geomagnetic storm forecast is: in spite of our growing knowledge we can predict only 75% of geomagnetic storms with *Dst* < - 80 *nT*, i.e. we can predict only 75% from 10% of the total number of magnetic storms, i.e. we can predict only 7.5% of all geomagnetic storms, i.e. we can predict almost nothing.

In reality prognostic quality is higher, because statistical laws and theory of probability do not allow accuracy level falling down lower than 40%. Geomagnetic storm prediction will remain on the probabilistically casual level of 40%-60% until focus of scientific community will be concentrated on investigation of severe and major storms' origin.

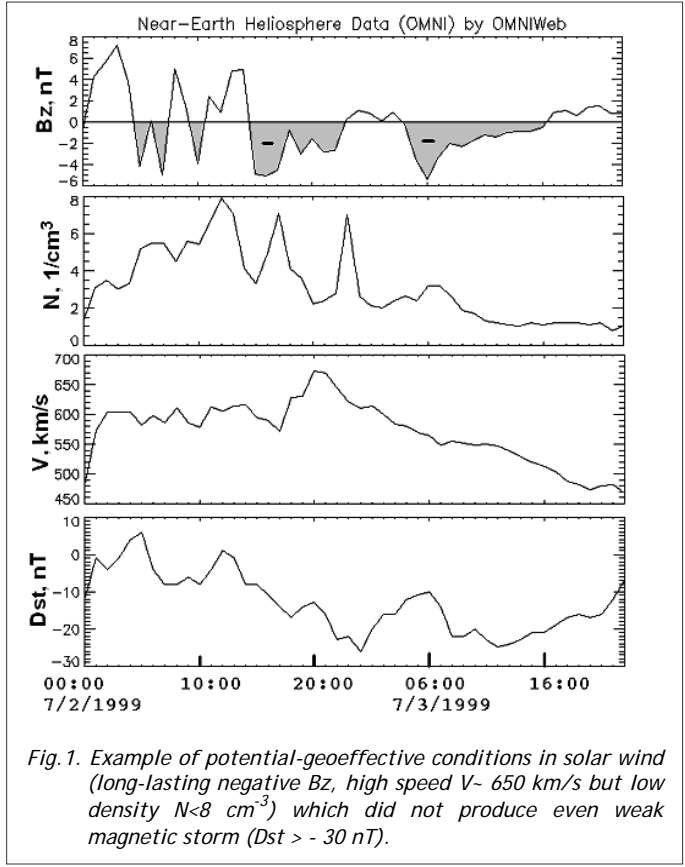

*Fig.1. Example of potential-geoeffective conditions in solar wind (long-lasting negative Bz, high speed V~ 650 km/s but low density N<8 $cm^{-3}$) which did not produce even weak magnetic storm (Dst > - 30 nT).*

Interest to less intense storms is not pure academic because moderate storms often produce much higher increases of relativistic electron fluxes near the geosynchronous orbit than intense storms do [7] and can lead to the satellite's anomalies and failures [8]. It was shown also most significant biological reaction is manly associated with weak and moderate storms [9].

The attempts to improve medium-term prognosis quality due to solar monitoring and estimation of ejecta probability or observation of coronal holes run across the problem of complicated propagation of different types of solar wind streams and their interactions. Most of weak and moderate magnetic storms are stimulated by streams of mixed origin, but calculation of appearance probability of mixed type of streams near the Earth's orbit is very difficult. So, more proper way is search for new geoeffective parameters.

Building of geomagnetic storm prognosis on the base of geoeffective structures recognition is a fruitful method, and its effectiveness may be improved due to investigations of solar wind conditions before and after onsets of magnetic storms of different intensities.

Geoeffectiveness of CME-like conditions of solar wind consists in a strong long-lasting southward IMF (stable negativity of IMF vertical $Bz$-component) and high velocity $V$, which start reconnection process on the dayside of magnetopause and fill magnetosphere by solar wind energy [3]. An explanation of high-speed geoeffective role is in its ability to provide more field lines for reconnection per time unit.

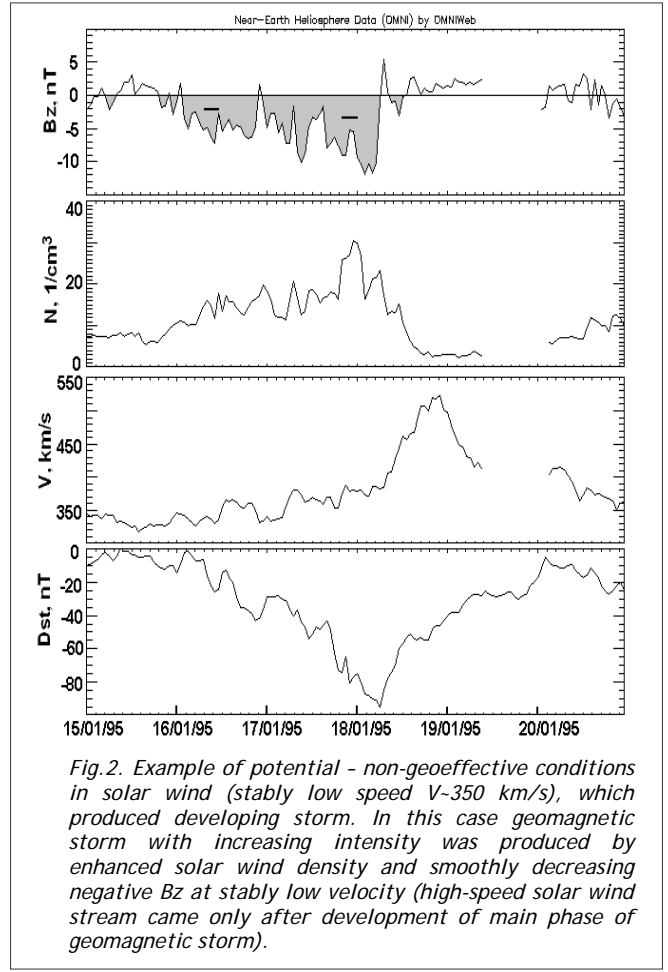

*Fig.2. Example of potential - non-geoeffective conditions in solar wind (stably low speed V~350 km/s), which produced developing storm. In this case geomagnetic storm with increasing intensity was produced by enhanced solar wind density and smoothly decreasing negative Bz at stably low velocity (high-speed solar wind stream came only after development of main phase of geomagnetic storm).*

But the question appears: "If we know that the most of severe storms obey this law (high $V$, strong and long-lasting $Bz$), whether these conditions are always associated with any geomagnetic storm?" The answer is "No".

It is known that only 23% of mild storms with - 50 $nT$ < $Dst$ < - 30 $nT$ are related to the high-velocity streams [2]. Fig.1 demonstrates the situation, when presence of long-lasting negative $Bz$ and high-speed stream was not enough for initiation of geomagnetic storm.

So, the investigations of conditions in solar wind, leading to the most intensive (but very rare) storms, do not throw light on the problem of origin of moderate and weak (but very often occurring) storms, and cannot help us to build the real-working medium-term prognosis of magnetic storms. Lows, found for intensive storms, can not be unconditionally expanded to the other types of geomagnetic storms, so it is necessary to reveal the rules of solar wind geoeffectiveness for weak and moderate storms exciting.

Orientation to the investigations of CME and high-speed streams with strong electric field leads to the situation, when other solar wind/IMF parameters (like plasma density $N$, level of turbulence, etc) mainly are not taken into account [2, 3, 10-14]. In particular, density is considered as a minor factor, just increasing the storm intensity or enhancing negative $Bz$ at the leading edge of magnetic cloud or inside corotating interaction regions (CIRs) [4, 13, 14].

Meanwhile, as it was statistically shown, the most of geomagnetic storms of years 1995 and 2000 were associated with increased and oscillated solar wind density (not with increased velocity) [15].

Case-study example is given in Fig. 2. It shows the situation, when simultaneously falling negative IMF $Bz$ and enhanced solar wind density without significant changes in velocity produce the geomagnetic storm with growing intensity.

Next figure (Fig.3) shows that geomagnetic storm may be result of non-simultaneous influence of high-density stream and negative $Bz$ at low solar wind velocity. It is interesting that this moderate storm happened in the "window", between two high-speed solar wind streams and there is time lag between the sharp increase of density and IMF turn to southward. Some other examples are given in [15].

Therefore CME-like conditions investigation is not panacea for successful medium-term prognosis, and additional investigations are needed.

## Possible ways of the problems' solving

As we see, the more proper way of future prognostic technique development lies through overcoming of main paradigm "CME-directed investigations" and changing direction to the medium-term prognosis of all magnetic storms (not only most intense storms). Possible ways of the problems' solving should be based on the tests of geoeffectiveness of poor-investigated features of solar wind.

For example, it was recently observationally revealed in [16, 17] that solar wind plasma density might enhance the geoeffectiveness of southward IMF and production of the ring current. It was shown in [18] that the quality of short-term storm prediction can be improved, especially for the most intense storms, by introduction into a forecasting algorithm of solar wind dynamic pressure. It was found that convective electric field oscillations of the order of $mHz$ are the good feature to warn about severe storm [19]. The new prognosis method, based on the SW density variations analysis before magnetic storms onset, was proposed in [20].

There are evidences that southward IMF conditions combined with high solar wind dynamic pressure immediately after a pressure front impact lead to enhanced coupling between the solar wind and the terrestrial magnetosphere, significantly increasing the geoeffectiveness of the solar wind [21].

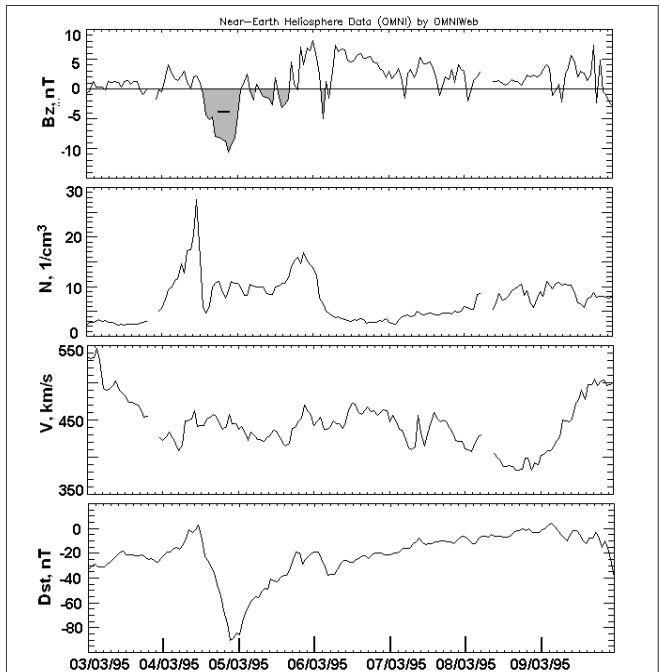

Fig. 3. Example of consequence of sharp solar wind density N increase and, then, deep fall of IMF Bz in low-speed solar wind, which produced moderate geomagnetic storm.

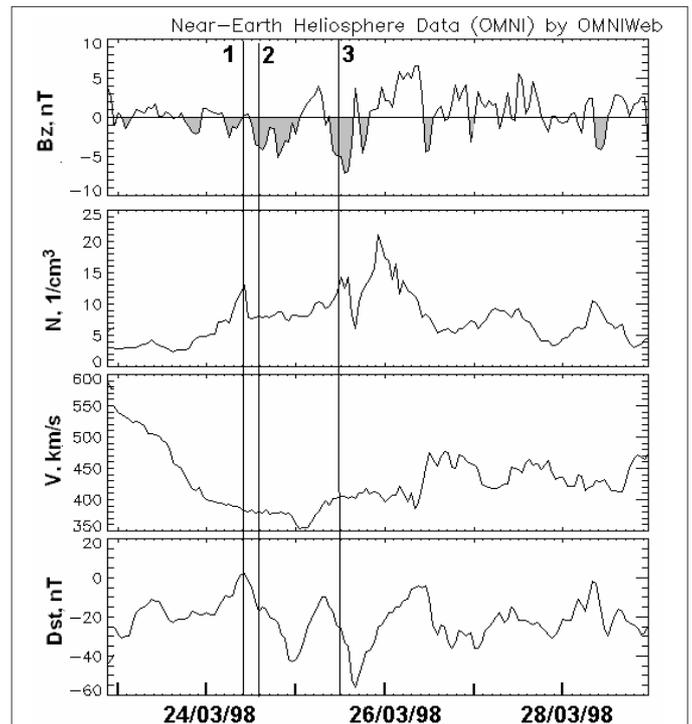

Fig. 4. Example of consequence of two geomagnetic storms, stimulated by dense southward solar wind with low speed. First one is a result of sharp solar wind density increase N~12 $1/cm^3$ and Bz ~ - 4 nT. Maximum density and moment of stable Bz turns to the negative values (and immediate beginning of weak storm) are marked correspondingly as 1 and 2. Second (moderate) geomagnetic storm is produced by repeated increase of solar wind density to ~ 14 $1/cm^3$ against a background of Bz ~ - 7 nT (the onset is marked as point 3).

Since solar wind dynamic pressure is practically completely controlled by solar wind density [15], the "increased density" factor in combination with negative IMF *Bz* (with lag or without lag) is the next pretender to strong geoeffective factor.

**TABLE 2**

| No | start | end | $N_{max}$, 1/cm$^3$ | min$Bz$, nT | Lag $dT$, hours | min$Dst$, nT | P |
|---|---|---|---|---|---|---|---|
| 1 | 15.01.1995 | 20.01.1995 | 20,7 | -14,3 | 17 | -95 | -14,3 |
| 2 | 03.03.1995 | 09.03.1995 | 27,5 | -8,8 | 9 | -90 | -8,8 |
| 3 | 25.03.1995 | 28.03.1995 | 55 | -15,4 | 1 | -107 | -15,4 |
| 4 | 21.04.1995 | 25.04.1995 | 27,5 | -7,9 | 3 | -53 | -7,9 |
| 5 | 30.11.1995 | 07.12.1995 | 17,7 | -9,2 | 1 | -62 | -9,2 |
| 6 | 23.05.1997 | 29.05.1997 | 32,4 | -10,4 | 10 | -73 | -10,4 |
| 7 | 16.09.1997 | 19.09.1997 | 12,6 | -9,1 | 8 | -56 | -9,1 |
| 8 | 12.11.1997 | 16.11.1997 | 24,7 | -6,4 | 15 | -49 | -6,4 |
| 9 | 29.12.1997 | 04.01.1998 | 34,7 | -10,4 | 8 | -77 | -10,4 |
| 10 | 15.02.1998 | 21.02.1998 | 12,4 | -15,1 | 7 | -100 | -15,1 |
| 11 | 23.03.1998 | 25.03.1998 | 11,6 | -5,3 | 12 | -43 | -5,3 |
| 12 | 25.03.1998 | 26.03.1998 | 14,3 | -7,2 | 1 | -56 | -7,2 |
| 13 | 04.08.1998 | 09.08.1998 | 23,2 | -19,3 | 1 | -138 | -19,3 |
| 14 | 16.08.1998 | 24.08.1998 | 25,6 | -10,2 | 3 | -67 | -10,2 |
| 15 | 16.10.1998 | 24.10.1998 | 65,3 | -16,7 | 10 | -112 | -16,7 |
| 16 | 11.11.1998 | 17.11.1998 | 33,4 | -17,6 | 6 | -128 | -17,6 |
| 17 | 08.12.1998 | 17.12.1998 | 11,2 | -12,7 | 7 | -69 | -12,7 |
| 18 | 28.02.1999 | 01.03.1999 | 22,6 | -13,4 | 9 | -94 | -13,4 |
| 19 | 01.03.1999 | 06.03.1999 | 56 | -14,4 | 4 | -95 | -14,4 |
| 20 | 14.11.1999 | 19.11.1999 | 19,1 | -11,5 | 2 | -79 | -11,5 |
| 21 | 20.01.2000 | 26.01.2000 | 28 | -15,7 | 7 | -97 | -15,7 |
| 22 | 29.03.2000 | 05.04.2000 | 18,7 | -7,2 | 3 | -60 | -7,2 |
| 23 | 25.10.2000 | 02.11.2000 | 39,3 | -17,1 | 3 | -127 | -17,1 |
| 24 | 17.12.2000 | 26.12.2000 | 24,9 | -13,9 | 5 | -62 | -13,9 |
| 25 | 24.02.2001 | 28.02.2001 | 44,6 | -6,4 | 4 | -37 | -6,4 |
| 26 | 20.04.2001 | 27.04.2001 | 29,7 | -12,8 | 14 | -102 | -12,8 |
| 27 | 09.09.2001 | 17.09.2001 | 21,4 | -9,7 | 35 | -57 | -9,7 |
| 28 | 30.10.2001 | 04.11.2001 | 23,4 | -12,5 | 4 | -106 | -12,5 |
| 29 | 10.12.2001 | 14.12.2001 | 26,6 | -6,1 | 3 | -39 | -6,1 |
| 30 | 27.12.2001 | 04.01.2002 | 67,2 | -9,8 | 14 | -58 | -9,8 |

Fig. 2 and 3 demonstrate very important feature of solar wind for geomagnetic storm triggering: storms may develop both after simultaneous density growth and IMF *Bz* turn to stable negative values and after consequence of these events. Sharp density growth must happen first. This is a necessary, but not a sufficient condition.

Negative *Bz* is the second rule, leading to geomagnetic storm onset. Apparently, density sharp growth turns the magnetosphere to the excited conditions and next negative IMF *Bz* allows to realize loading-unloading mechanism in the magnetosphere.

Fig.4 is an additional example of consequence of two geomagnetic storms, obeying to the rule "sharp solar wind density increase + negative IMF *Bz* = weak or moderate geomagnetic storm".

Case-study investigation of 30 most indicative geomagnetic storms, stimulated by increased density of solar wind and negative IMF *Bz* with a lag at non-significant changes of low velocity, shows that geomagnetic storm may start even if N-*Bz* time delay is about several hours (see Table 2).

Table 2 includes tested time intervals with events in solar wind, leading to storms of different intensities. Start and end days of the intervals are given (not storm onsets and ends!). Maximum value of solar wind density Nmax before storm onset, minimum IMF *Bz* (*minBz*) value during geomagnetic storm, time lag *dT* between density maximum and *Bz* minimum, *Dst* minimum during a storm (it is necessary to remark that *Bz* minimum and *Dst* minimum also have a time lag), and values of fitting parameter P:

$$P = \min Bz - \sqrt{N_{max} dT} \qquad (1)$$

are presented in Table 2.

It was found that minimum *Dst* values during such type of storms might be calculated from *P* as follow:

$$\min Dst = -4.5 + 6.5P \qquad (2)$$

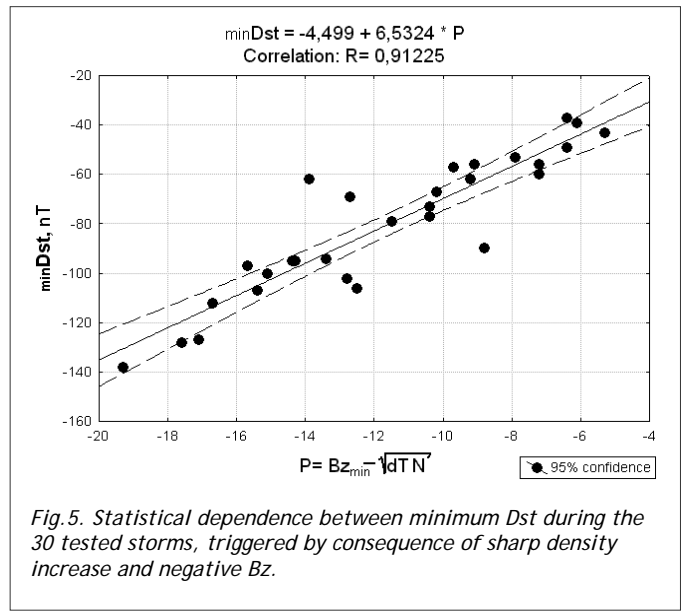

*Fig.5. Statistical dependence between minimum Dst during the 30 tested storms, triggered by consequence of sharp density increase and negative Bz.*

Fig.5 shows the correlation between these parameters (correlation coefficient equals 0.91), i.e. storm intensity strongly depends on previous behavior of *Bz*, *N* and time lag between them.

Therefore even preliminary investigations give the key to the best understanding of solar wind geoeffectiveness function and clear up more significant density role that it was assume before.

Some statistical results, obtained recently by Khabarova et al [15, 22], concerning the search for new prognostic factors and estimation of solar wind geoeffectiveness, confirm the density taking into account importance:

1. The solar wind behavior before and after the onsets of all magnetic storms is different from the well-known

behavior of the solar wind before and after severe magnetic storms. Statistical analysis of geomagnetic storms with storm sudden commencement (SSC) for 40 years and all types storms for 15 years allows to suggest that the well-known rule: "High speed + long-lasting negative $Bz$ + compression = geomagnetic storm" does not work for most geomagnetic storms.

Pre- and after-storm solar wind features are different from the statistical average of solar wind conditions [22]:

- solar wind density, Alfvén Mach number and plasma beta are higher and velocity is lower in the neighborhood of magnetic storm onsets;

- most geoeffective solar wind streams flow from regions located upper Earth orbit plane.

2. Test on geoeffectiveness of corotating interaction regions (CIRs) and magnetic clouds (MCs) during both the solar minimum and maximum (1995-96 and 2000-01) shows, that CIRs and MCs have nearly equal input in the production of medium and severe magnetic storms [15], in a good correspondence with Yermolaev et al [23] results, but in controversy with commonly accepted point of view about prevailing geoeffectiveness of MCs.

3. Statistical relationships between the main solar wind and IMF parameters ($VB$, $VBz$, $Dst$, $V$, $Bz$, $N$, $Kp$) have turned out to differ at various solar cycle phases. This fact may indicate that intrinsic properties of the solar wind and IMF, as well as their magnetospheric response, vary during a solar cycle [15], and prediction algorithms must adapt to these variations, otherwise they would be not equally effective during various phases of solar cycle.

The important result of correlative analysis is that the correlation between $N$ and $Bz$ is practically absent. Thus, the hypothesis about an increase of southward IMF by an enhanced $N$ [14] must be called in question. Meanwhile the correlation of $N$ with IMF magnitude $B$ is much higher. Thus, the solar wind density indeed can drag and compress the IMF lines, but $N$ equally enhances IMF of any direction, not only southward.

4. Case-study analysis and test of solar density fluctuations level in $ULF$ range show that the solar wind behavior before a magnetic storm persistently demonstrates important features. Besides the rapid increase of the plasma density, provoking magnetic storm beginning, a more gradual increase of $N$ occurs for a few hours or even days before the main density growth. The increase of $N$ is not steady, but is accompanied by irregular fluctuations [15].

## Discussion and Conclusions

The main problems of medium-term magnetic storm forecasting are effect of the shift of scientific interest to prognosis of severe magnetic storms only and to estimation of probability of CME registration near the Earth's orbit. The most hopeful way of their solving is searching for additional prognostic factors in solar wind.

The recent works show that variations and sharp changes of the solar wind plasma and IMF are a largely underestimated factor in magnetic storm triggering and could be effectively used for space weather forecasting analysis.

The studies show that the solar wind density plays a more significant geoeffective role than it was previously assumed. A sharp density increase and consequent negative $Bz$ can produce weak, moderate and even strong magnetic storms without any significant changes of the solar wind velocity. So, the well-known rule: "High speed + long-lasting negative $Bz$ + compression = severe geomagnetic storm" must be supplemented with the rule for weak and moderate geomagnetic storms: "sharp solar wind density increase + negative IMF $Bz$ = weak or moderate geomagnetic storm".

It is possible to explain the second rule by prevailing of "loading-unloading mechanism" of magnetospheric energy transfer from the solar wind in the most of cases of mixed type solar wind streams' interaction with magnetosphere over the "directly driven magnetosphere" mechanism, which is more appropriate for explanation of solar wind – magnetosphere interaction during the streams like ICME's and CIR's crossing the Earth's orbit.

The debates about these two types of mechanisms have been hold for years, and assumption that they both may be realized in the magnetosphere and provide different types of geomagnetic storms is useful for practical aims (i.e. for medium-term forecasting) [24, 25].

The triggering role of density is not investigated properly until now and the rule "density + negative $Bz$" is not revealed clearly with standard statistical analysis because the time delay between the sharp increase of $N$, $Bz$ minimum and $Dst$ minimum varies substantially from storm to storm.

The recent results show necessity of prognostic usage not only magnetic field disturbance level, but also density fluctuations level.

**Acknowledgment**
This work was partly supported by INTAS grant # 05-1000008-8050.